\documentclass[aps,preprint]{revtex4}%
\usepackage{amsfonts}
\usepackage{amsmath}
\usepackage{amssymb}
\usepackage{float} 
\usepackage{subfig}
\usepackage{graphicx}%
\providecommand{\U}[1]{\protect \rule{.1in}{.1in}}

\begin{document}
\title{\textbf{Exact solution of two fluid plasma equations
for the creation of jet-like flows and seed magnetic fields in
cylindrical geometry} }
\date{\today}
\author{H. Saleem$^{1,2}$ \\
$^{1}$Department of Space Science, Institute of Space\\
Technology (IST), Islamabad, Pakistan\\
$^{2}$Theoretical Research Institute Pakistan Academy of Sciences (TRIPAS),\\
Islamabad, Pakistan\\
Email: saleemhpk@hotmail.com}

\begin{abstract}
An exact solution of two fluid ideal classical plasma equations is
presented which shows that the jet-like outflow and magnetic field
are generated simultaneously by the density and temperature
gradients of both electrons and ions. Particular profiles of density
function $\psi=\ln \bar{n}$ (where $\bar{n}$ is normalized by some
constant density $N_0$) and temperatures $T_j$ (for $j=e,i)$ are
chosen which reduce the set of nonlinear partial differential
equations to two simple linear equations generating longitudinally
uniform axial outflow and magnetic field in cylindrical geometry in
several astrophysical objects. This mechanism also seems to be
operative for producing short scale plasma jets in the solar
atmosphere in the form of spicules and flares. The presented
solution requires particular profiles of density and temperatures,
but it is a natural solution of the two fluid ideal classical plasma
equations. Similar jet-like outflows can be generated by the density
and temperature gradients in neutral fluids as well.

\newpage

\end{abstract}
\maketitle

\bigskip

\section{1. INTRODUCTION}

First observation of astrophysical jets was made long ago \cite{cur}
and later observations showed that jets are collimated outflows of
gases and plasmas from a variety of sources ranging from young
stellar objects (YSOs) to active galactic nuclei (AGNs) \cite{jen,
mes,fer, bel}. Jets emerging from YSOs have speeds $v$ smaller than
speed of light $c$ $(v\leq10^{-3} c)$ and hence can be treated using
classical plasma models while the jets emerging from AGNs have
speeds approaching speed of light and are highly relativistic . AGN
jets have sizes of the order of $10^{6} pc$ and YSO jets have sizes
of the order of $(10^{-4}-1) pc$ . In between these two extremes are
the outflows of gases, electron ion plasmas and electron positron
plasmas from normal stars, neutron stars, massive X-ray binary
systems, galactic stellar mass black holes (or microquasars).

Magnetic fields are considered to be an important constituent of
astrophysical jets emerging from accretion disks, compact stars and
black holes \cite{bla, con, fuk}. Jets of active galactic nuclei
(AGN) are extended up to kilo or mega parsecs with helical magnetic
fields \cite{chr, gab1, gab2}. The magnetic fields of these objects
are generated by large scale currents and are amplified by dynamo
effects. Gamma ray bursts are also associated with relativistic
plasma outflow. Solar corona has millisions of plasma jets
(spicules) and also flares \cite{pri, sal1, ste, haw}. The formation
of short scale solar spicules and flares seems to be very similar to
large scale non-relativistic galactic jets. Cylindrical structures
of plasma also emerge in solar corona along with loops and play a
role in solar coronal heating.

Theoretical models and numerical studies have been performed to
understand the dynamics of relativistic and non-relativistic jets
\cite{shi, ouy,sto, nak}. Non-relativistic simulations for the
outflows of plasma from accretion disks have been performed using
magnetohydrodynamic (MHD) equations \cite{ust, hon}. Simulations
using neutral fluid equations have been performed to compare
dynamics of relativistic and non-relativistic jets \cite{ros}.
Kelvin-Helmholtz instabilities have also been investigated in
supersonic jets in cartesian and cylinrical geometries \cite{mic}.
Plasma jets are also produced in laboratories to understand the
physics of jet formation \cite{cia, loe}.

Despite a large difference in velocities, scale sizes, luminosities
and magnetic energies of the jets emerging from different sources,
the physical origin is believed to be very similar. Many physical
mechanisms take part in the formation, stability and acceleration of
astrophysical jets including rotation, Lorentz force, magnetic field
topology, dissipative forces, gravitation and pressure gradients.
Extremely large gravitational effects in case of material outflow
from massive black holes require general relativity to be invoked.
Flows of hot ionized material create currents and Lorentz force.

Observations reveal that in YSOs, some jets rotate in the opposite
direction to the rotation of the source i.e. star or disk.
Theoretical model has been presented to explain the counter-rotation
of astrophysical jets and winds using classical magnetohydrodynamics
(MHD) \cite{sau}. They have shown that the counter rotation is
induced by variation of velocity along the flow, or by shocks, or by
inhomogeneous magnetic field. They have verified the observed
counter-rotation speed both by analytical approach and by computer
simulation.

The fundamental source of energy is the thermal energy which is
converted into flow and electromagnetic energies in these systems
particularly in jets moving with velocities much smaller than speed
of light. Our aim is to find out a simple physical mechanism which
generates perpendicular jet-like outflows in non-relativistic
plasmas and neutral fluids. This means the fundamental source for
astrophysical jets is based on thermodynamics coupled with fluid
dynamics.

Biermann effect produced by electron baroclinic vector is believed
to be the source of seed magnetic fields in stars \cite{bie} and
galaxies \cite{laz}. It has been pointed out that it is not only the
electron baro clinic vector which produces seed magnetic field;
rather the ion dynamics is also coupled with the electron dynamics
in this mechanism. Hence ions should not be treated as a background
of stationary positive charge \cite{sal2,sal3}. In these studies,
cartesian geometry was used and main focus was on the generation of
seed magnetic field, therefore an additional assumption was made
that magnetic field ${\bf B}$ is related with the vorticity $(\nabla
\times {\bf v})$ of plasma as ${\bf B}=\alpha (\nabla \times {\bf
v})$ where $\alpha$ was a constant. A theoretical model for the
generation of magnetic field in relativistic plasma of early
universe has also been presented \cite{mah}.

The ideal classical MHD equations were solved by decomposing the
velocity vector into components in cylindrical geometry to explain
counter rotation of jets \cite{sau}. Then these authors showed that
the rotation velocity component can change sign if the velocity of
flow along the flux tube becomes smaller than a threshold value. On
the other hand, here it is demonstrated that if plasma density and
temperatures have particular profiles, then the axial longitudinally
uniform flow ${\bf v}$ and magnetic field ${\bf B}$ are created
simultaneously. The flow becomes function of density and increases
with time. Dissipative terms and gravity are ignored to highlight
the physical mechanism and to find out an exact analytical solution
of nonlinear partial differential equations. Similarly, it is shown
that jet-like flows can also be created in neutral fluids by the
density and temperature gradients.

\section{Two Fluid Plasma Jets}
Here it is shown that thermodynamic forces produced by density and
temperature gradients of electron and ion fluids create jet-like
plasma flows along z-axis in cylindrical geometry. Only electron
dynamics is not sufficient to produce such flows; rather it is found
that dynamics of ions is coupled with the electrons and hence both
positive and negative fluids generate seed magnetic field and plasma
flow simultaneously. For the investigation of this long time
behavior, electron inertia is ignored $(m_e \rightarrow 0)$ and
hence momentum conservation yields,
\begin{equation}
0\simeq -e n_e \left({\bf E}+\frac{1}{c} {\bf v}_e \times {\bf
B}\right)-\nabla p_e \label{1}
\end{equation}

In the limit $\mid\partial_t\mid<<\omega_{pe},\Omega_e,\mid
c\nabla\mid$ where $\omega_{pe}=(\frac{4\pi n_e
e^2}{m_e})^{\frac{1}{2}}$ is the electron plasma oscillation
frequency and $\Omega_e=(\frac{eB}{m_e c})$ is the electron gyro
frequency, Maxwell's equation reduces to Amperes' law
\begin{equation}
\nabla\times {\bf B}=\frac{4 \pi}{c}{\bf j} \label{2}
\end{equation}
and quasi-neutrality $(n_e\simeq n_i=n)$ holds. Electron velocity
can be defined as,
\begin{equation}
{\bf v}_e={\bf v}_i - \frac{c}{4 \pi e} \left(\frac{\nabla\times{\bf
B}}{n}\right) \label{3}
\end{equation}
Curl of Eq. (1) yields
\begin{equation}
-\frac{1}{c}\partial_t {\bf B}=-\frac{1}{c}\nabla\times({\bf v}_e
\times {\bf B})-\nabla\times(\frac{\nabla p_e}{en})\label{4}
\end{equation}
Using (3) and (4), we obtain

\begin{equation}
\partial_t {\bf B}=\nabla \times ({\bf v}_i \times
{\bf B})-(\frac{c}{4\pi n e}) \{\nabla \times [(\nabla \times {\bf
B})\times {\bf B}]\}\label{5}
\end{equation}
$$
+\frac{c}{4 \pi n e} \{\nabla \psi\times [(\nabla \times {\bf
B})\times {\bf B}]\}-\frac{c}{e} (\nabla\psi \times \nabla T_e)
$$
where $\psi=\ln \bar{n}$, $\bar{n}=\frac{n}{N_0}$ and $N_0$ is
arbitrary constant density. Curl of ion equation of motion
\begin{equation}
(\partial_t+{\bf v}_i\cdot\nabla){\bf v}_i=en\left({\bf
E}+\frac{1}{c}{\bf v}_i \times {\bf B} \right) -\nabla p_i \label{6}
\end{equation}
yields,

\begin{equation}
a \partial_t {\bf B}+\partial_t(\nabla \times{\bf v}_i)=\nabla
\times [{\bf v}_i\times(\nabla \times{\bf v}_i)]+a\nabla\times({\bf
v}_i\times{\bf B})+\frac{1}{m_i}(\nabla \psi \times \nabla
T_i)\label{7}
\end{equation}
where $a=\frac{e}{m_i c}$. Ideal gas law $p_j=n_j T_j$ has been used
for both electron and ion fluids where $j=e,i$. Let us assume
longitudinally uniform flow by imposing the condition $\nabla \cdot
{\bf v}_j=0$ and hence density does not vary with time i.e.
$\partial_t n=0$. The continuity equations
\begin{equation}
\partial_t n +n \nabla \cdot {\bf v}_j +\nabla n \cdot {\bf v}_j=0
\label{8}
\end{equation}
demand
\begin{equation}
\nabla \psi \cdot {\bf v}_j=0 \label{9}
\end{equation}
which due to (3) also requires,
\begin{equation}
\nabla \psi \cdot (\nabla \times {\bf B})=0 \label{10}
\end{equation}

Let the plasma density vary in $(r,\theta)$-plane and temperatures
be functions of $z$ coordinate only, viz,
\begin{equation}
\psi=\psi(r,\theta) \label{11}
\end{equation}
and
\begin{equation}
T_j=-T_j^{\prime} z =-(\frac{T_{j0}}{\delta_j})z\label{12}
\end{equation}
where $T_j^{\prime}$ is constant and it denotes derivative of
$T_{j}$ with respect to $z$ and $\delta_j$ is the scale length of
the temperature gradient. Equations (11) and (12) yield,
\begin{equation}
\nabla \psi=(\partial_z \psi){\bf e}_r+(\frac{1}{r}\partial_{\theta}
\psi){\bf e}_{\theta}\label{13}
\end{equation}
and
\begin{equation}
\nabla T_j=-\frac{T_j}{\delta_j} {\bf e}_z\label{14}
\end{equation}
Here ${\bf e}_r$,${\bf e}_z$, and ${\bf e}_{\theta}$ are unit
vectors along radial, axial and angular directions, respectively.
Note that when $\psi=0$, we have $n=N_0$. Using equations (11-14),
we find
\begin{equation}
(\nabla \psi \times \nabla
T_e)=(-\frac{T_{e0}}{r}\partial_{\theta}\psi){\bf e}_r +
(\frac{T_{e0}}{\delta_e}\partial_r \psi){\bf e}_{\theta} \label{15}
\end{equation}
and
\begin{equation}
(\nabla \psi \times \nabla
T_i)=(-\frac{T_{i0}}{r}\partial_{\theta}\psi){\bf e}_r +
(\frac{T_{i0}}{\delta_i}\partial_r \psi){\bf e}_{\theta} \label{16}
\end{equation}
Gradient of $\psi$ in $(r,\theta)$-plane along with the condition
(10) forces ${\bf B}$ to have the following form of spatial
dependence,
\begin{equation}
{\bf B}=(\nabla \chi(r,\theta) \times {\bf e}_z)=\frac{1}{r}
(\partial_{\theta} \chi ){\bf e}_r+(-\partial_r \chi) {\bf
e}_{\theta} \label{17}
\end{equation}
which yields,
\begin{equation}
\nabla \times {\bf B}=-\nabla^2 \chi(r,\theta){\bf e}_z\label{18}
\end{equation}
Note that the vector potential ${\bf A}$ has only $z$-component
non-zero i.e. ${\bf A}=A_z {\bf e}_z=\chi {\bf e}_z$ and hence the
magnetic helicity $\emph{H}$ is zero, viz,
\begin{equation}
\emph{H}={\bf A}\cdot {\bf B}=0 \label{19}
\end{equation}

Our aim is to show that time-independent forms of density and
temperature gradients are able to create time-dependent jet-like
flows in the axial direction perpendicular to $(r,\theta)$-plane.
Now we prove that following form of plasma axial flow is created by
the above mentioned profiles of $\nabla \psi$ and $\nabla T_j$,

\begin{equation}
{\bf v}_i=u(r,\theta) f(t) {\bf e}_z \label{20}
\end{equation}

It will be seen later that plasma flow in axial direction becomes a
function of density and temperature. The nonlinear terms of electron
and ion fluid equations can be expressed in terms of scalars defined
above as  follows,

\begin{equation}
\nabla \times ({\bf v}_i \times {\bf B})=\{u,\chi\} f(t) {\bf e}_z
\label{21}
\end{equation}

\begin{equation}
\nabla \times [(\nabla \times {\bf B})\times {\bf
B}]=\{\chi,\nabla^2 \chi \} {\bf e}_z \label{22}
\end{equation}

and
\begin{equation}
\nabla \psi \times [(\nabla \times {\bf B})\times {\bf
B}]=[\{\chi,\psi \}\nabla^2\chi] {\bf e}_z \label{23}
\end{equation}
where
\begin{equation}
\{f,g\}=\frac{1}{r}(\partial_r f
\partial_{\theta}g-\partial_{\theta}f, \partial_r g)\label{24}
\end{equation}

Note that

\begin{equation}
\nabla \times [{\bf v}_i \times (\nabla \times {\bf
v}_i)]=0\label{25}
\end{equation}
Let us impose the following conditions on density, magnetic field
and flow,
\begin{equation}
\{\chi,\psi \}=\{u,\chi \}=0=\{\chi,\nabla^2 \chi \}=0 \label{26}
\end{equation}
Interestingly under these conditions all the nonlinear terms of
Eqs.(5) and (7) vanish and they reduce, respectively, to
\begin{equation}
\partial_t {\bf B}=-\frac{c}{e} (\nabla \psi \times \nabla
T_e)\label{27}
\end{equation}
and
\begin{equation}
a\partial_t {\bf B}+\partial_t(\nabla \times {\bf
v}_i)=\frac{1}{m_i}(\nabla \psi \times \nabla T_i)\label{28}
\end{equation}
Due to (20), we have
\begin{equation}
\nabla\times {\bf v}_i=\{(\frac{1}{r}\partial_{\theta} u){\bf
e}_r-(\partial_r u){\bf e}_{\theta}\}f(t) \label{29}
\end{equation}
therefore ion vorticity (29), baroclinic vectors (15,16) and
magnetic field (17) all become parallel to each other. Using (20)
and (13) in equations (27) and (28), we obtain
\begin{equation}
\partial_t\chi=\frac{cT_{e0}^{\prime}}{e} \psi \label{30}
\end{equation}
and
\begin{equation}
u\partial_t f=-(\frac{T_{e0}^{\prime}+T_{i0}^{\prime}}{m_i})\psi
\label{31}
\end{equation}
Let $\chi=h(r,\theta) f(t)$ and integrate (30) and (31) for
$t:0\rightarrow \tau$ to get, respectively,

\begin{equation}
\chi(r,\theta,\tau)=h(r,\theta)f(\tau)=\frac{cT_{e0}^{\prime}}{e}
\psi(r,\theta) \tau\label{32}
\end{equation}
and
\begin{equation}
v_i(r,\theta,\tau)=u(r,\theta)f(\tau)=
-(\frac{T_{e0}^{\prime}+T_{i0}^{\prime}}{m_i})\psi(r,\theta) \tau
\label{33}
\end{equation}
Since $u=u(\psi)$ and $\chi=\chi(\psi)$, therefore $\{\chi,\psi\}=0$
and $\{u,\chi\}=0$ hold. Note that $f(t)=(constant) t$ has been
assumed.

Now we try to find out an exact solution of the nonlinear two fluid
plasma equations by choosing an appropriate form of $\psi$ which
fulfills all the above assumptions. For this, we impose a condition
on $\psi$, such that
\begin{equation}
\nabla^2 \psi=-\lambda^2 \psi \label{34}
\end{equation}
where $\lambda$ is a constant. Following relations of Bessel
functions hold,
\begin{equation}
\frac{d J_1 (\lambda r)}{dr}=\lambda J_0(\lambda r) -\frac{1}{r}
J_1(\lambda r) \label{35}
\end{equation}
and
\begin{equation}
\frac{d^2 J_1 (\lambda r)}{d r^2}=-(\frac{2 \lambda^2}{r^2})J_1
(\lambda r) -\frac{\lambda}{r} J_0(\lambda r) \label {36}
\end{equation}
where $J_0$ and $J_1$ are Bessel functions of order zero and one,
respectively. The magnetic field $\chi$ and plasma flow $u$ have
been expressed in terms of density function $\psi$ in equations (30)
and (31). Let us define,
\begin{equation}
\psi(r,\theta)=\psi_0 J_1(\lambda r) \cos (\theta) \label{37}
\end{equation}
where $\psi_0$ is the magnitude of $\psi$ and $\lambda$ is the scale
length of density variation. With this definition of $\psi$ the
condition $ \{\chi,\nabla^2 \chi\}=0$ mentioned in (26) also holds.
Density function $\psi$ approaches zero for large $\lambda r$. The
form of this function is illustrated in Fig. (1). Flow and magnetic
field have also similar forms but the direction of flow is axial and
it increases with time. Since all physical quantities vary like
$J_1$ therefore this solution is valid in the limited region of the
source where $\psi$ has the given form. At large $\lambda r$ the
density becomes equal to the background density, say $n_0$ and
phenomenon disappears.

\begin{figure}[H]
	\hspace {4.5cm}
	\includegraphics[width = 2.5in, height = 2.0in]{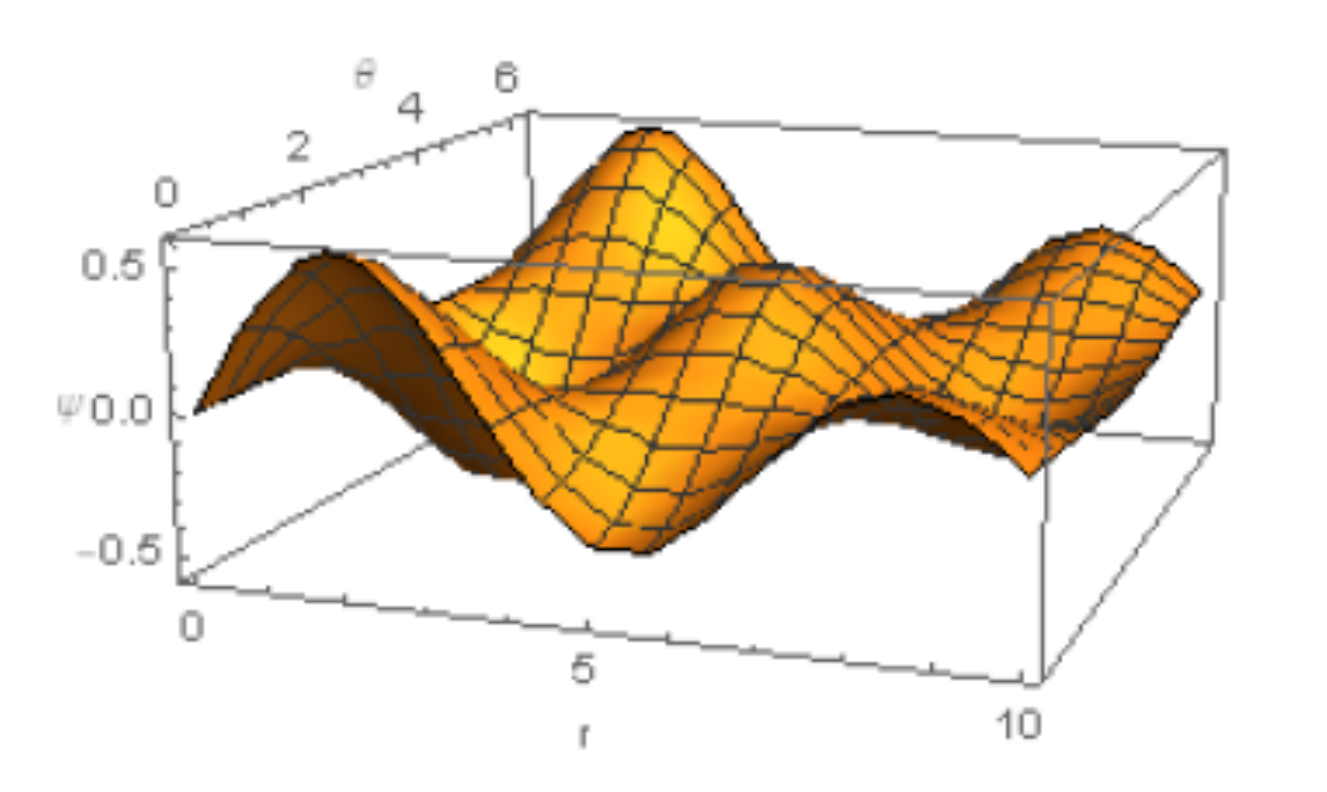} 
	\caption{Density function $\psi=\ln \bar{n}$ is plotted vs
		$\lambda r:0\rightarrow 10$ and $\theta:0\rightarrow 2\pi$. Similar
		axial flow ${\bf v}$ out of plasma disc is created and it increases
		with time because it is directly proportional to $\psi$ and $t$.}
\end{figure}

It is very interesting to note that the form of density function
given in (37) fulfills all the above conditions. Thus an exact
solution of two fluid plasma equations (5) and (7) exists in
cylindrical geometry which predicts that both magnetic field and ion
vorticity are generated simultaneously by the baro clinic vectors of
electrons and ions. Plasma is ejected in the z-direction like a jet
perpendicular to $(r, \theta)$ plane. Velocity $u$ becomes zero
where $J_1=0$ or $\cos \theta =0$. Strictly speaking these are the
points where flow disappears. This means the flow under this
theoretical model is broken in pieces. At these points the density
becomes equal to background density.

\section{Neutral Fluid Jets}
In this section, we show that neutral fluid is also ejected out of a
two dimensional material spread in $(r,\theta)$ plane in the form of
a jet along $z$-axis in the cylindrical geometry. Let us consider an
ideal neutral fluid with density $\rho$, velocity ${\bf v}$ and
pressure $p$. Its dynamics is governed by the following two simple
equations,
\begin{equation}
\rho(\partial_t+{\bf v} \cdot \nabla){\bf v}=- \nabla p \label{38}
\end{equation}
and
\begin{equation}
\partial_t n+ \nabla \cdot (n {\bf v})=0 \label{39}
\end{equation}
where $\rho= m n$ and $m$ is mass of the fluid particle. We again
assume longitudinally uniform flow $\nabla \cdot {\bf v}=0$ with
time independent density i.e. $\partial_t \rho=0$. Then the
continuity equation demands $\nabla n \cdot {\bf v}=0$ which can be
expressed as
\begin{equation}
\nabla \psi \cdot {\bf v}=0 \label{40}
\end{equation}
where $\psi=\ln{\bar{n}}$ and $\bar{n}=\frac{n}{N_0}$ and $N_0$ is
constant density at axis of the cylinder $(r=0)$. Curl of equation
(38) gives,
\begin{equation}
\partial_t(\nabla \times {\bf v})=\nabla\times [{\bf
v}\times (\nabla\times{\bf v})]+\frac{1}{m} (\nabla \psi \times
\nabla T) \label{41}
\end{equation}
If nonlinear term vanishes
\begin{equation}
\nabla \times \{{\bf v} \times (\nabla \times {\bf v})\}=0
\label{42}
\end{equation}
then we obtain a simpler equation for the vorticity generation
\begin{equation}
\partial_t(\nabla \times {\bf v})=\frac{1}{m} (\nabla \psi \times
\nabla T) \label{43}
\end{equation}
This shows that vorticity can be generated in the neutral fluid by
the baro clinic vector. If this is true, then we need to find out a
suitable solution which satisfies the conditions (40) and (42) along
with equation (43). Let us consider density and temperature profiles
as follows,
\begin{equation}
\psi=\psi(r,\theta) \label{44}
\end{equation}
and
\begin{equation}
T=-T_0^{\prime}z=-\frac{T}{\delta}z \label{45}
\end{equation}
which give,
\begin{equation}
\nabla \psi=(\partial_t \psi){\bf
e}_r+(\frac{1}{r}\partial_{\theta}\psi){\bf e}_{\theta} \label{46}
\end{equation}
and
\begin{equation}
\nabla T=-T_0^{\prime}{\bf e}_z=-\frac{T}{\delta}{\bf e}_z
\label{47}
\end{equation}
where $T_0^{\prime}=\frac{T_0}{\delta_0}$ is a constant, $\delta_0$
is temperature gradient scale length and $\hat{z}$ is unit vector
along axis of the cylinder. Equations (46) and (47) yield,
\begin{equation}
(\nabla\psi \times \nabla
T)=-T_0^{\prime}\left[(\frac{1}{r}\partial_{\theta} \psi){\bf
e}_r+(-\partial_r \psi){\bf e}_{\theta}\right]\label{48}
\end{equation}
Equation (43) shows that vorticity is parallel to $(\nabla \psi
\times \nabla T)$, hence we choose the form of flow as,
\begin{equation}
{\bf v}=u(r,\theta) f(t) {\bf e}_z \label{49}
\end{equation}
Then (43) can be expressed as,
\begin{equation}
\{(\frac{1}{r}\partial_{\theta} u){\bf e}_r+(-\partial_r u){\bf
e}_{\theta}\}\partial_t
f=-\frac{T_0^{\prime}}{m}\{(\frac{1}{r}\partial_{\theta} \psi){\bf
e}_r+(-\partial_r \psi){\bf e}_{\theta}\}\label{50}
\end{equation}
Integrating for $t=0\rightarrow \tau$, it yields,

\begin{equation}
\{(\frac{1}{r}\partial_{\theta} u){\bf e}_r+(-\partial_r u){\bf
e}_{\theta}\}\partial_t
f=-\frac{T_0^{\prime}}{m}\{(\frac{1}{r}\partial_{\theta} \psi){\bf
e}_r+(-\partial_r \psi){\bf e}_{\theta}\}\tau \label{51}
\end{equation}
Equation (50) gives a relation between flow $u$ and density $\psi$
as
\begin{equation}
u(r,\theta)=-\frac{T_0^{\prime}}{m} \psi(r,\theta) \label{52}
\end{equation}
where
\begin{equation}
f(t)=(constant)t\label{53}
\end{equation}
Now we try to choose a particular form of $\psi$ which should
satisfy all the above mentioned assumptions. Let
\begin{equation}
\psi=\psi_0 J_1(\lambda r) \cos \theta \label{54}
\end{equation}
 The density at $r=0$ is constant $N_0$ and hence
${\bf v}=0$ at the axis. However, vorticity $(\nabla\times {\bf v})$
will have several zeros in $(r,\theta)$-plane because this plane is
expanded along $r$. Then we find
\begin{equation}
u(r,\theta)=-\frac{T_0^{\prime}}{m} \psi_0 J_1(\mu r) \cos \theta
\label{55}
\end{equation}
Equation (52) shows that neutral material is ejected in the form of
jet along $z$-axis out of two dimensional density structure while
the temperature gradient is along the direction of flow.
\section{Discussion}
Keeping in view the long time plasma behavior, a formulism has been
presented in which all the physical quantities turn out to be the
functions of density $\psi$ chosen in Eq. (37). Interestingly, all
the nonlinear terms vanish naturally and hence all the assumptions
used to derive the simple linear equations (27) and (28) are
satisfied. Solution of these equations under the condition (34) is
given by Eq.(37). Here $\psi_0$ is the amplitude and $\psi$ varies
along radial direction like Bessel function of order one $(J_1)$ and
in $\theta$-direction the variation is chosen to be like $\cos
\theta$. This form of density function reduces the set of nonlinear
partial differential equations to two linear equations. It seems
important to mention here that the presented solution requires
particular profiles of density and temperatures, but it is a natural
solution of the two fluid classical plasma equations which predicts
the creation of jet-like plasma outflow ${\bf v}_{(r,\theta)}$ and
the magnetic field ${\bf B}_{(r,\theta)}$ simultaneously. Unlike
self-gravitating systems \cite{ben}, we just focus attention on the
investigation of outflows from a plasma and a fluid source (star or
a disc) and do not include additional forces like gravity.

First we have considered a two fluid classical ideal plasma to show
that the driving force for the observed astrophysical and laboratory
jets is based on the structures of density and temperature
gradients. This theoretical model suggests that magnetic field and
plasma outflow are produced simultaneously in astrophysical objects
and they evolve with time from a non-equilibrium $(T_e\neq T_i)$ two
fluid ideal classical plasma. Important point to note is that the
highly nonlinear system of two fluid plasma equations can create an
ordered structure in the form of a jet. The driving force of such
flows and associated magnetic fields is produced by the
thermodynamic free energy available in the form of density and
temperature gradients. Similar mechanism in classical plasma in the
Sun's atmosphere seems to be operative for producing short scale
plasma jets in millions (the spicules) and solar flares as well as
cylindrical plasma structures in solar corona.

Second, it has also been shown that jet-like outflows are also
created in neutral fluids by the same physical mechanism. Here it
has also been found that the baro clinic vector of a neutral fluid
produces jet like structures expelling the material in the axial
direction. Thus density and temperature gradients seem to be the
fundamental source of jet formation in both neutral fluids and
classical plasma systems.

The Fig. (1) illustrates the profile of $\psi$ for the range of
$\lambda r: 0\rightarrow 10$ and $\theta:0\rightarrow 2\pi$.
Equation (33) gives the form of axial flow as a function of $\psi$
but increasing with time because flow is directly proportional to
time $t$.  Similar flow can be created in neutral fluids as well.
Thus axial flows can be created by density and temperature gradients
in both neutral fluids and plasmas.

\begin{acknowledgments}
Author is very grateful to Professor Zensho Yoshida of Tokyo
University for reading the initial draft of this manuscript and
giving several very valuable suggestions.
\end{acknowledgments}

\end{document}